\documentclass[]{aa}   
\usepackage{subfigure}
\usepackage[varg]{txfonts}                
\usepackage{graphicx}
\graphicspath{{/}}                          
\usepackage{amssymb}
\usepackage{natbib,twoopt}
\usepackage{comment}
\usepackage{multirow}
\bibpunct{(}{)}{;}{a}{}{,} 
\makeatletter

\renewcommand{\arraystretch}{1.5}

\makeatother

\begin{document}

\title{Changes in the cyclotron line energy on short and long timescales in V~0332+53}

\author{V. Vybornov\inst{1,2}
\and V. Doroshenko\inst{1}
\and R. Staubert \inst{1} 
\and A. Santangelo\inst{1}}

\offprints{\email{vybornov@astro.uni-tuebingen.de}}

\institute{Institut f\"ur Astronomie und Astrophysik, Universit\"at T\"ubingen, 
Sand 1, 72076 T\"ubingen, Germany
 \and Space Research Institute, 
 Profsouznaya 84/32, 117997 Moscow, Russia}


\abstract{We present the results of the pulse-amplitude-resolved spectroscopy
of the accreting pulsar V~0332+53 using the NuSTAR observations of the
source in 2015 and 2016. We investigate the dependence of the energy of the
cyclotron resonant scattering feature (CRSF) as a function of X-ray luminosity
on  timescales comparable with the spin period of the pulsar within individual
observations, and the 
behavior on longer timescales within and between the two observed outbursts. 
We confirm that in both cases the CRSF energy is negatively
correlated with flux at luminosities higher than the critical luminosity and is
positively correlated at lower luminosities. We also confirm the recently
reported gradual decrease in the line energy during the giant outburst in 2015.
Using the NuSTAR data, we find that this decrease was consistent with a
linear decay throughout most of the outburst, and flattened or even reversed at
the end of the 2015 outburst, approximately simultaneously with the transition
to the  subcritical regime. We also confirm that by the following outburst in 2016 the line 
energy rebounded to previous values. The observed behavior of the
CRSF energy with time is discussed in terms of changes in the geometry of the
CRSF forming region caused by changes in the effective magnetospheric radius.}

\maketitle

\section{Introduction}
\label{sec:intro}
 Cyclotron resonance scattering features (CRSFs) are observed in X-ray spectra
of some accreting pulsars\footnote{http://www.iasfbo.inaf.it/\textasciitilde mauro/pulsar\_list.html}
\citep{1978ApJ...219L.105T,2004AIPC..714..323H,2005A&A...433L..45K}.
The observed energy of these line-like features $E_{\mathrm{cyc}}$,
called cyclotron lines,  is believed to be directly related to the electrons gyro-frequency 
and thus to the strength of the magnetic field in the line-forming region as 
$E_{\mathrm{cyc}}$ = $11.6[\textrm{keV}]B_{12}n/(z+1)$, 
where $B_{12}$ is the magnetic field strength in units of $10^{12}$ gauss, 
$n$ is the Landau level number, and $z$ is the gravitational redshift at the line-forming region.
The energy of the cyclotron line was found to change with the rotational phase
of the neutron star, luminosity, and time, which can potentially  probe the
geometry of the line-forming region.

The Be-transient X-ray pulsar V~0332+53, about 7\,kpc  from the Sun, is a
unique cyclotron line source that exhibits a clear negative correlation of the
observed CRSF energy with X-ray luminosity $L_{\textrm{x}}$ at high luminosities \citep{2006MNRAS.371...19T}. 
This behavior was
associated with the appearance and growth of the radiatively supported
extended emission region above the surface of the  neutron star, known as
the accretion column \citep{1976MNRAS.175..395B,
2006MNRAS.371...19T,2013ApJ...777..115P}. 

On the other hand, sources that show the opposite behavior, i.e., a
positive correlation, are also known: Her X-1
\citep{2007A&A...465L..25S}, A~0535+26 \citep{2011A&A...532A.126K}, Vela X-1
\citep{2014ApJ...780..133F, 2016MNRAS.463..185L}, GX~304-1
\citep{2011PASJ...63S.751Y,2015A&A...581A.121M,2017MNRAS.466.2752R}, and Cep X-4
\citep{2017A&A...601A.126V}. 
Several interpretations have been proposed
to explain the observed positive $E_{\textrm{cyc}}/L_{\textrm{x}}$
correlations in these sources. The CRSF is believed to form closer to the
surface of the neutron star with the observed energy related either to the vertical
distribution of plasma emissivity \citep{2017MNRAS.466.2752R,2017A&A...601A.126V} 
or to the velocity \citep{2015MNRAS.454.2714M} within the line-forming region. In the first case, the increase in the accretion rate is assumed to
effectively reduce the height of the emission region, thus shifting the CRSF to
higher energies. In the alternative scenario, the observed line energy is
red-shifted due to the bulk motion in the accretion flow by a factor proportional
to the velocity in the line-forming  region. An increase in the accretion rate implies
lower velocities and redshift, and so the CRSF also moves to higher energies.
The observed positive $E_{\mathrm{cyc}}$/$L_{\textrm{x}}$ correlation
is thus associated with the absence of an extended accretion column in both cases. 
One could expect, therefore, that a transition between positive and negative correlation 
will be observed at the luminosity corresponding to the onset of an accretion column.

Indeed, a transition from the negative correlation to the positive correlation of the cyclotron 
line in V~0332+53 below $L_\mathrm{x}\sim10^{37}$\,erg\,s$^{-1}$ has been recently reported in 
\cite{2017MNRAS.466.2143D} based on the analysis of NuSTAR
data obtained during the 2015 giant outburst. 

In addition to the luminosity dependence, V~0332+53 also exhibited a very peculiar
gradual decay of the observed cyclotron line energy with time during the 2015 outburst
\citep{2016MNRAS.460L..99C}. 
The decay is attributed either to an accretion-induced decrease in the neutron star's intrinsic magnetic field \citep{2016MNRAS.460L..99C} 
or to changes in the emission region geometry associated with the changes in the accretion disk 
structure \citep{2017MNRAS.466.2143D}.

Regardless of the interpretation, the observed ``hysteresis'' of $E_{\textrm{cyc}}$ due to the time decay 
complicates the analysis of the relation between line energy and X-ray luminosity, in
particular of the observed transition from the negative correlation to the positive  at
low luminosities. On short timescales, however, variations with time can be neglected. 

Here we report on the results of the analysis
of the $E_{\mathrm{cyc}}$ behavior during the two consecutive outbursts in 2015 and 2016, 
both on long and short timescales using a  ``pulse-to-pulse'' technique or
``pulse-amplitude-resolved'' analysis \citep{2011A&A...532A.126K}.
The primary goal of this investigation is to verify the results reported previously for the
long-term behavior of the CRSF \citep{2016MNRAS.460L..99C,2017MNRAS.466.2143D}
through comparison with the pulse-to-pulse technique.

\section{Observations and spectral analysis}
\label{sec:obs}
\begin{figure}[h]
\center{\includegraphics[width=1\linewidth]{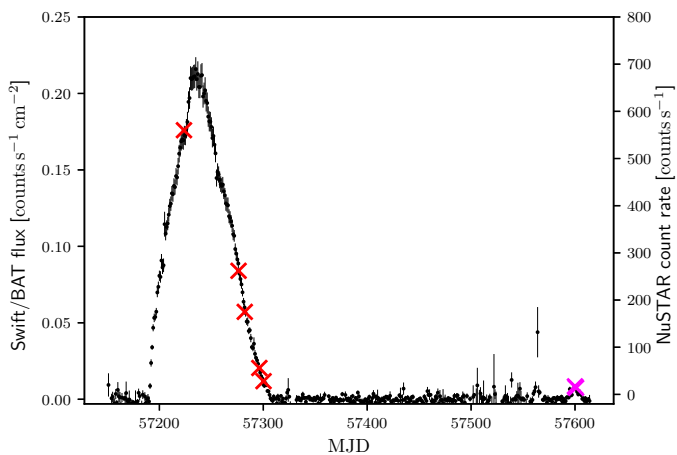}}
\caption{NuSTAR pointed observations of V~0332+53 during the 2015 giant outburst (red crosses) 
and the 2016 minor outburst (magenta crosses);   the Swift/BAT light curve in the 15-50 keV energy band 
are shown as black data points.}
\label{fig:BAT_NuSTAR}
\end{figure}

In the present work we analyze the data of the NuSTAR observations of V0332+53 during the 2015 giant and 
the 2016 minor outburst (Fig.~\ref{fig:BAT_NuSTAR}), referred to below as the 2015 and 2016 outburst. 
Although we use the BAT light curve\footnote{https://swift.gsfc.nasa.gov/results/transients/V0332p53/} to show 
where the NuSTAR observations are, we do not analyze BAT data as 
in \cite{2017MNRAS.466.2143D} to avoid uncertainties related to the cross-calibration.
A short summary of the NuSTAR observations is given in Table~\ref{table:obs_params}. 

The data extraction was performed using the \texttt{nupipeline} and
\texttt{nuproducts} utilities distributed as part of HEASoft 6.19 (CALDB
20170120). Source spectra and light curves were extracted from a circular
region with a radius of 80$^{\prime\prime}$ centered on the source. The
background was extracted from a circular region with a radius of
100$^{\prime\prime}$ situated as far as possible from the source. The analysis
of the extracted spectra was carried out using XSPEC 12.9 and Sherpa CIAO 4.9
packages. 

\begin{figure}[h]
\center{\includegraphics[width=1\linewidth]{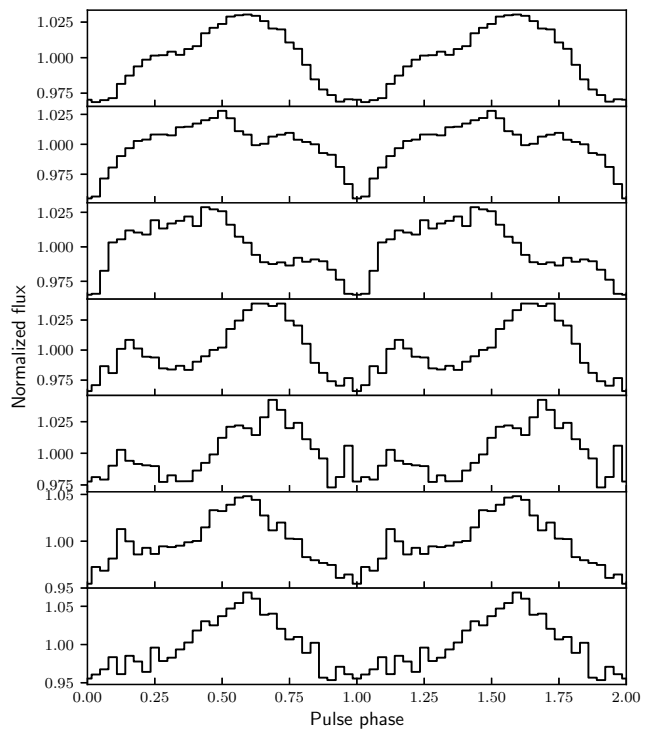}}
\caption{NuSTAR pulse profiles in the 3-79 keV energy range. 
The first five pulse profiles from the top correspond to the 2015 giant outburst, whereas the last two profiles belong to the 
2016 minor outburst.}
\label{fig:pprofs}
\end{figure}

In addition to the fundamental line at around 30\,keV, the first harmonic of the CRSF at $\sim$~50\,keV is also  
detected in the broadband spectrum of the source. 
Since it is quite prominent, the first harmonic can also affect the fundamental line parameters; 
the size of the extraction regions was chosen to optimize the signal-to-noise ratio
at high energies where the harmonic is detected.

The spectral analysis consisted of two stages, namely, a standard analysis of integrated spectra and 
a pulse-amplitude-resolved analysis.
All spectra were analyzed in the 5-78 keV energy range using the 
\texttt{compTT} model \citep{1994ApJ...434..570T} as a continuum and a multiplicative Gaussian line of the form
\begin{equation}
G(E) = 1 - d e^{-\ln 2\frac{(E-E_{\mathrm{cyc}})^2}{\sigma_{\mathrm{cyc}}^2}}
\end{equation}
to describe the CRSFs.  
Here $d$ is the depth, $\sigma_{\mathrm{cyc}}$ is the
width, and $E_{\mathrm{cyc}}$ is the centroid energy of the line. 
The shape selection for the CRSFs is based on a slightly better description of these
features in comparison with the exponential Gaussian line
(\texttt{gabs}\footnote{https://heasarc.gsfc.nasa.gov/xanadu/xspec/manual/node23
2.html} in XSPEC). Further details on  the model selection can be
found in \cite{2017MNRAS.466.2143D}.
In addition, this model selection allows us to compare and control 
parameters of integrated spectra with \cite{2017MNRAS.466.2143D} because we analyzed spectra extracted from the same 
NuSTAR observations, although in a slightly different manner. For instance, we included the harmonic of the CRSF, but
despite the optimization of the extraction region, the width of the harmonic is
poorly constrained owing to lower statistics,
in particular in the weak observations.
Thus, we fixed the width of the harmonic to a value of 10 keV
found for the brightest observation where the parameters of the line are well constrained.
The Fe $K_\alpha$ line is also  not detected significantly at low fluxes, and so  we
fixed the energy and width of the line to values measured during the brightest
observation. Despite a slightly different energy range and the absence of the BAT data, 
the continuum and the fundamental cyclotron line parameters are very close to the values obtained 
by \cite{2017MNRAS.466.2143D}, which is not surprising and so we do not list them here.

\begin{table}
\caption{Parameters of the analyzed NuSTAR observations.}
\label{table:obs_params}
\centering
{\renewcommand{\arraystretch}{1.2}
\setlength{\tabcolsep}{3pt}
\begin{tabular}{ | c | c c c c |}
\hline
& ObsID & Date (MJD)  & Exp. (ks) & Period (s) \\
\hline
\multirow{6}{*}{\rotatebox[origin=c]{90}{2015}} & & Group 1 & & \\
& 80102002002 & 57223.42-57223.85 & 10.5 & 4.3761(6) \\
& 80102002004 & 57275.94-57276.42 & 14.9 & 4.3759(0) \\
& 80102002006 & 57281.92-57282.36 & 17.0 & 4.3758(9) \\
& & Group 2 & & \\
& 80102002008 & 57295.96-57296.40 & 18.1 & 4.3759(4) \\
& 80102002010 & 57299.99-57300.49 & 20.8 & 4.3759(6) \\
\hline
\multirow{2}{*}{\rotatebox[origin=c]{90}{2016}} & 90202031002 & 57599.75-57600.27 &  25.2 & 4.3762(8) \\
& 90202031004 & 57600.76-57601.27 &  25.0 & 4.3763(5) \\
\hline
\end{tabular}
}
\end{table}

To assess the variation in the cyclotron line energy with flux within individual
observations, we produced pulse-amplitude-resolved spectra by applying the
 pulse-to-pulse technique elaborated in \cite{2011A&A...532A.126K}. 
This method uses flux variability that a source exhibits from pulse to pulse, allowing us to 
investigate spectral-luminosity dependencies on timescales on the order of the pulse period.

\begin{figure}[h]
\center{\includegraphics[width=1\linewidth]{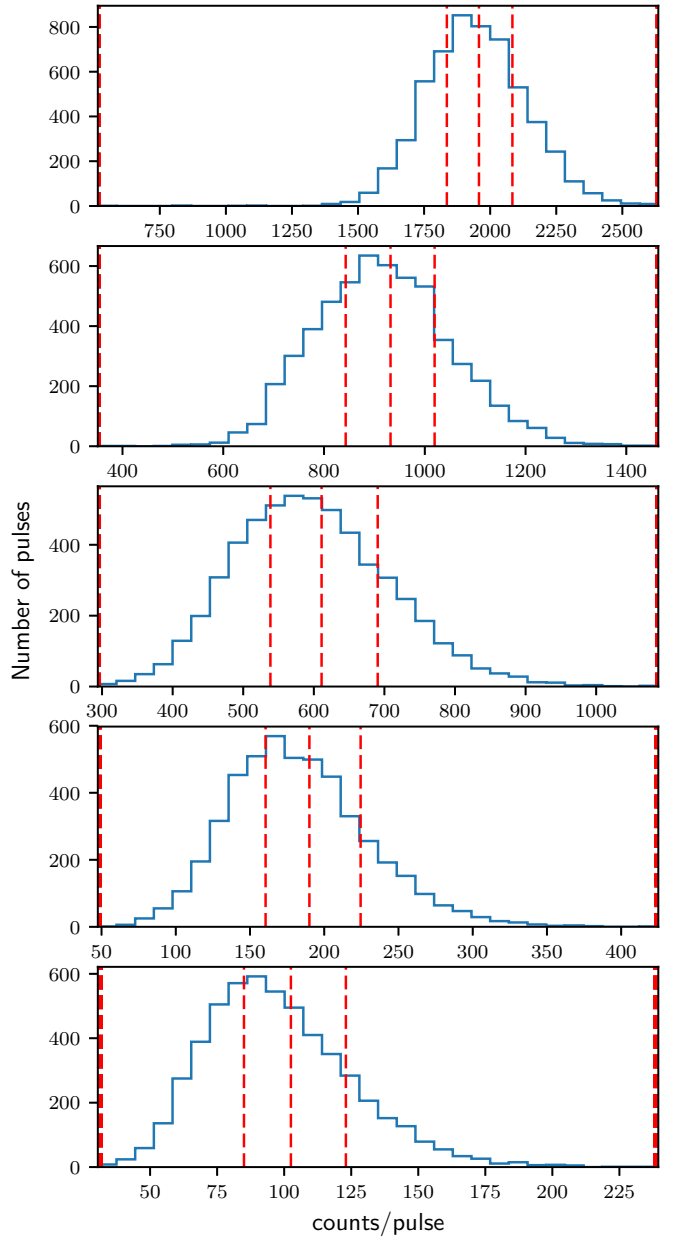}}
\caption{Normalized distributions of counts in an individual pulse for the observations during the giant outburst used in
the pulse-amplitude-resolved analysis. The date of observations increases from top to bottom. The red dashed lines 
indicate the boundaries of the amplitude bins in which the total number of counts is equal.}
\label{fig:pulse_distribs}
\end{figure}

First of all, using the binary- and barycenter-corrected data, we measured the pulse period of the source using the NuSTAR observations 
(Table~\ref{table:obs_params}) and produced the pulse profiles  shown in Fig.~\ref{fig:pprofs}. 
For every pulse we measured a number of counts in the whole pulse (in 3 - 78 keV). 
The number of counts is equivalent to the pulse amplitude 
and is thus used as a measure of the pulse brightness. We then built a frequency distribution of pulses as a function of 
the total number of counts in a pulse (Fig.~\ref{fig:pulse_distribs}). We verified that the distribution is wider 
than the Poisson distribution with the same mean value of counts in a pulse, otherwise the observed pulse amplitude 
variability is caused by statistical fluctuations. 
Based on the observed distribution of pulse amplitudes we then divided all the
pulses into four groups to conduct the pulse-amplitude-resolved spectral analysis.
The number of groups is somewhat arbitrary and depends on the available
counting statistics, so that it was defined experimentally
to ensure that the cyclotron line centroid energy is well constrained when fitting the resulting spectra.
We derived a pulse using good time intervals (GTIs), then defined to which part of the distribution it belongs by calculating 
the number of counts in this pulse, and finally stacked up the counts of the all pulses that fall into the same part of the 
distribution.
This procedure was performed for five NuSTAR observations of the 2015 giant outburst. 
The observations of the 2016 outburst have insufficient statistics to apply the pulse-to-pulse technique.

We also took  into account the dead-time and verified that the selected intervals are statistically
independent, i.e., the mean number of photons detected per cycle is significantly
different for each interval. The spectra of the source and the background of each interval were
extracted and described using the same model as described above. In the case of limited
counting statistics, we had to fix the continuum parameters, the width of the
fundamental, and parameters of the harmonic to values obtained for
the flux-averaged spectrum, concentrating thus only on the changes in the fundamental centroid energy.

\begin{figure}[h]
\center{\includegraphics[width=1\linewidth]{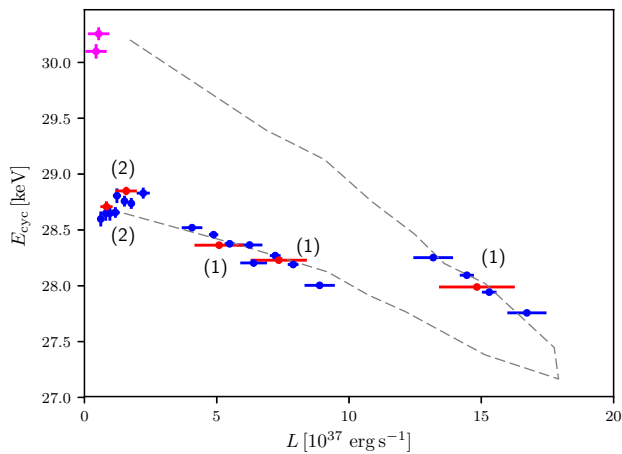}}
\caption{Dependence of the energy of the fundamental cyclotron line on luminosity. The two positive and three 
negative correlations clearly seen in the figure are derived from the pulse-to-pulse analysis (blue crosses). 
The red and magenta crosses are values obtained from the pulse-averaged spectra of the 2015 giant outburst 
and the 2016 outburst. 
The observations with negative p2p-correlations comprise Group~1, while Group~2 consists of two observations with 
positive p2p-correlations, labeled (1) and (2), respectively.
The gray line reflects the ``hysteresis'' behavior observed with Swift/BAT \citep{2016MNRAS.460L..99C}.}
\label{fig:corrs}
\end{figure}

\begin{figure}[h]
\center{\includegraphics[width=1\linewidth]{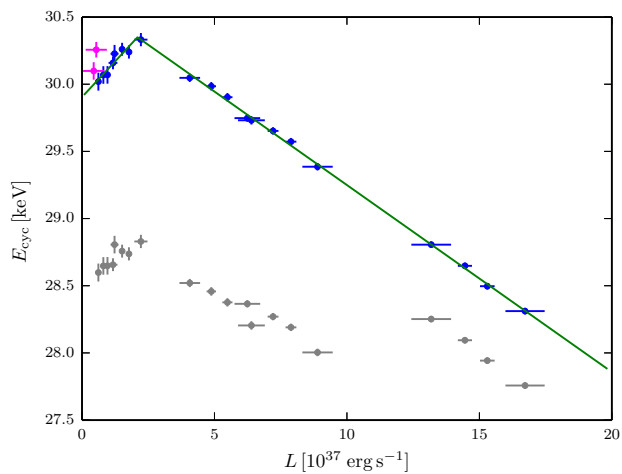}}
\caption{Observed correlations between the fundamental cyclotron line energy and luminosity as a result of 
the pulse-amplitude-resolved analysis (gray crosses).
The blue crosses are the correlations corrected for the time drift. 
The green line is the best fit model to the blue crosses, which reflect the $E_{\textrm{cyc}}/L_{\textrm{x}}$ dependence corrected for the time drift.
The reference time corresponds to the beginning of the giant outburst. 
The data of the 2016 outburst are marked as the magenta crosses.}
\label{fig:corrs_fit}
\end{figure}

\section{Results}
\label{sec:results}

The analysis results in several $E_{\mathrm{cyc}}$/$L_{\mathrm{x}}$ 
correlations obtained on short timescales for each observation and presented in Fig.~\ref{fig:corrs}, 
which we refer to below as ``{p2p}'' correlations.
In particular, Fig.~\ref{fig:corrs} shows that there are two distinct groups of observations, i.e.,
showing a positive or negative correlation of the line energy with flux.

Group~1 includes the observations carried out at high luminosities  
(ObsID 80102002002, 80102002004, and 80102002006;  see Table 1), 
where a negative $E_{\mathrm{cyc}}$/$L_{\textrm{x}}$ correlation is observed. 
For Group~2 (ObsID 80102002008 and 80102002010) the trend reverses to a clear positive 
$E_{\mathrm{cyc}}$/$L_{\textrm{x}}$ correlation, confirming the transition already reported 
by \cite{2017MNRAS.466.2143D}
Here, however, we can measure the variations in  the line energy more precisely because we get a greater number of 
data points with sufficiently small error bars splitting each observation according to pulse amplitude. This is very important for 
the low-flux observations, for which we obtained eight points instead of two.

\begin{table}
\caption{Luminosities and cyclotron line energies of the best-fit models of spectra produced 
by the pulse-to-pulse technique. The luminosities are derived based on observed fluxes in the 5-78 keV energy range 
assuming the source distance of 7 kpc; uncertainties for luminosities are correspond to 1$\sigma$-uncertanties in bins of the count
distributions represented in Fig.~\ref{fig:pulse_distribs}. Other model parameters were fixed to values obtained from averaged
spectra.}
\label{table:p2p_params}
\centering
{\renewcommand{\arraystretch}{1.4}
\setlength{\tabcolsep}{3pt}
\begin{tabular}{| c | c c c |}
\hline
ObsID & $L_{5-78 \,\mathrm{keV}}$ & $E_{\mathrm{cyc}}$ & $\chi^2/\mathrm{d.o.f}$ \\
 & ($10^{37}$ erg/s) & (keV) & \\
\hline
\multirow{4}{*}{\rotatebox[origin=c]{90}{80102002002}} & $16.7\pm0.7$ & $27.76\pm 0.02$ & \multirow{4}{*}{1.055/5565} \\
& $15.3\pm0.3$ & $27.94\pm 0.02$ & \\
& $14.5\pm0.3$ & $28.09\pm 0.02$ & \\
& $13.2\pm0.8$ & $28.25\pm 0.02$ & \\
\hline
\multirow{4}{*}{\rotatebox[origin=c]{90}{80102002004}} & $8.9\pm0.6$ & $28.00\pm 0.03$ & \multirow{4}{*}{1.061/5227}  \\
& $7.9\pm0.2$ & $28.19\pm 0.03$ & \\
& $7.2\pm0.2$ & $28.27\pm 0.03$ & \\
& $6.2\pm0.5$ & $28.37\pm 0.03$ & \\
\hline
\multirow{4}{*}{\rotatebox[origin=c]{90}{80102002006}} & $6.4\pm 0.5$ & $28.20\pm 0.03$ & \multirow{4}{*}{1.074/5027} \\
& $5.5\pm0.2$ & $28.38\pm 0.03$ & \\
& $4.9\pm0.2$ & $28.46\pm 0.03$ & \\
& $4.1\pm0.4$ & $28.52\pm 0.03$ & \\
\hline
\multirow{4}{*}{\rotatebox[origin=c]{90}{80102002008}} & $2.2\pm 0.2$ & $28.83\pm 0.05$ & \multirow{4}{*}{1.036/4132}  \\
& $1.8\pm0.1$ & $28.73\pm 0.05$ & \\
& $1.5\pm0.1$ & $28.76\pm 0.05$ & \\
& $1.2\pm0.2$ & $28.66\pm 0.05$ & \\
\hline
\multirow{4}{*}{\rotatebox[origin=c]{90}{80102002010}} & $1.2\pm 0.2$ & $28.83\pm 0.05$ &  \multirow{4}{*}{1.033/3849} \\
& $0.96\pm 0.05$ & $28.73\pm 0.05$ & \\
& $0.81\pm 0.04$ & $28.76\pm 0.05$ & \\
& $0.62\pm0.09$ & $28.66\pm 0.05$ & \\
\hline
\end{tabular}
}
\end{table}

The dependence of $E_{\mathrm{cyc}}$ on luminosity turns out to be linear and the 
slopes are consistent with being the same for all observations within the two groups. 
This becomes evident when the time dependence of $E_{\mathrm{cyc}}$ is removed (see Fig.~\ref{fig:corrs_fit}).
The shift between the individual observations is associated with the previously reported  
decrease in  cyclotron energy with time \citep{2016MNRAS.460L..99C}. 
Assuming that each observation probes an ``instantaneous'' behavior of the cyclotron line 
(the influence of the time variation is negligible within individual observations), 
we estimated the rate of the decrease in the line energy using the observed shifts based on \emph{NuSTAR} data alone. 
So, fitting all five {p2p}-correlations simultaneously using a broken
linear model, similarly to \cite{2017MNRAS.466.2143D}, assuming common slopes within each group,
is successful and results in the representation in the upper part of Fig.~\ref{fig:corrs_fit}.
The broken linear model has the following free parameters: the luminosity at which
the correlation changes from positive to negative, the corresponding value for $E_{\mathrm{cyc}}$
(the reference time is the onset of the 2015 outburst,  MJD~57190.0), the two slopes of the
linear correlations, and the value for the (assumed linear) drift of $E_{\mathrm{cyc}}$ with time.

We find the slopes for the two groups of $-0.139 \pm 0.007$ keV/[$10^{37} \mathrm{erg} \,\mathrm{s}^{-1}$] 
and $0.21 \pm 0.08$ keV/[$10^{37} \mathrm{erg} \,\mathrm{s}^{-1}$] and, therefore, we confirm the reported 
transition from a negative correlation at high luminosities to a positive
correlation at low luminosities in V~0332+53 with high significance ($\sim22\sigma$). 
The transition luminosity is found to be $(2.1 \pm  0.4) \times 10^{37}$ erg/s 
(assumed at MJD~$\sim$~57288). The decay rate of $E_{\mathrm{cyc}}$ in the 
observations with negative correlation is found to be $-0.0162 \pm 0.0009$ keV/day, 
which is consistent with the values reported by \cite{2017MNRAS.466.2143D}. 
The total decrease in line energy over the entire outburst is then $\sim 1.5$ keV, 
consistent with the value reported by \cite{2016MNRAS.460L..99C}.

We note that after Her~X-1, for which the first combined analysis of the coexisting luminosity and 
time dependences of a cyclotron line was reported  \citep{2016A&A...590A..91S}, and Vela~X-1 
\citep{2016MNRAS.463..185L}, V~0332+53 is now the third source that shows such behavior. 

Figure~\ref{fig:corrs_fit} shows data points corrected for the time variation (blue crosses) and for the combined fit
reflecting the behavior of the CRSF energy with luminosity in the giant outburst corrected for a linear decay in time.
Consistent with the report in \cite{2017MNRAS.466.2143D}, the line energies measured in the 2016 outburst 
(magenta crosses) do not seem  to be shifted, that is the drop of the CRSF energy during the giant outburst was
completely recovered by that time.

It is the first time that both the pulse-amplitude-resolved and the traditional 
pulse-amplitude-averaged analyses of the same observations of an outburst of the same source 
give the same result, demonstrating the potential of this novel technique.

\begin{figure}[h]
\center{\includegraphics[width=1\linewidth]{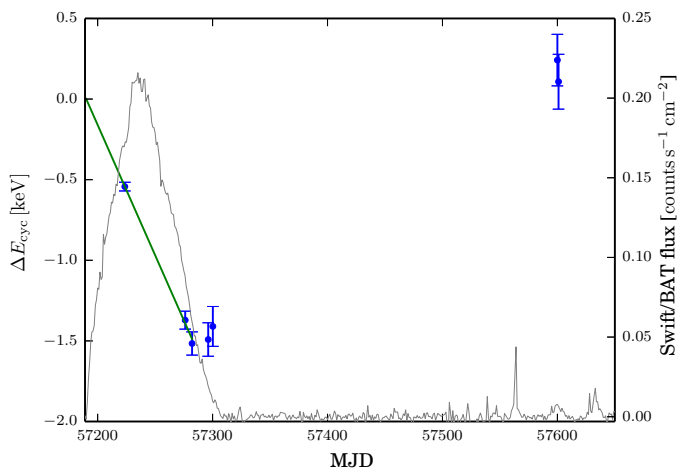}}
\caption{Energy offsets (blue bars) due to the time drift modeled as a broken linear function (green line). 
The offsets are measured from the beginning of the giant outburst ($T_0 = 57190.0$ MJD). 
The gray line is the 15-50 keV Swift/BAT time profile similar to that in Fig.~\ref{fig:BAT_NuSTAR}.}
\label{fig:drift}
\end{figure}

It also is interesting to note that the overall decay of the cyclotron line energy seems to
slow down or even reverse closer to the end of the outburst, as shown in Fig.~\ref{fig:drift}. 
If we describe observed line energy offsets with a broken linear function (i.e., flattening to a constant after a break), 
the break surprisingly occurs at MJD $\sim$ 57282(4),  close to the transition between the two accretion 
regimes (MJD $\sim$ 57288), which might suggest that there is a connection between these two events. 
The significance of the break in the linear drift of the cyclotron energy can be estimated, 
for instance, using the multiple linear regression (MLR) test \citep{2002ApJ...571..545P}. 
The linear fit to the 2016 data gives $\chi^2\sim9$ for three degrees of freedom, whereas the statistics improve 
to $\chi^2\sim0.63$ for two degrees of freedom using the broken linear model. 
The more complex model is thus preferred at $\sim99.6$\% confidence level based on the MLR test. 
Alternatively, one can assume that the break time coincides with the moment of the accretion regime transition and 
estimate the slopes before and after the break, which turn out to be -0.016(1) and 0.02(3), respectively, 
i.e., they differ at $\sim90$\% confidence level.

\section{Discussion}
\label{sec:disc}
A possible transition from a negative to a positive $E_{\textrm{cyc}}/L_{\textrm{x}}$ correlation has been reported by 
\cite{2017MNRAS.466.2143D} based on the analysis of averaged NuSTAR and Swift/BAT spectra.
This transition likely reflects the change in the accretion regime from the super- to subcritical.
This conclusion, however, is hampered to some extent by a generally complex evolution of the line energy throughout
the 2015 outburst because of a peculiar gradual decay 
and by uncertainties in the energy cross-calibration of Swift/BAT 
and NuSTAR and a low number of observations below the transition.
As a result, the statistical significance of the observed transition in \cite{2017MNRAS.466.2143D} was fairly low.

To confirm or refute the presence of the transition, we carefully analyzed the evolution of the cyclotron line during the 2015 and 
2016 outbursts  by applying the pulse-amplitude-resolved spectral analysis using NuSTAR data alone. 
Using data of a single instrument we avoid uncertainties related to cross-calibration issues. This approach 
also allows us to decouple the time dependence of the line energy from changes related to luminosity by
obtaining $E_{\textrm{cyc}}/L_{\textrm{x}}$ correlations measured on short timescales. As shown above, 
the slopes of the short timescale correlations were found to be the same as those measured on long timescales, 
i.e., the long-term decay of the line and flux-related variations are indeed independent.

Our analysis also revealed that the linear decay of the line energy reported in \cite{2017MNRAS.466.2143D}
seems to flatten out at later stages of the 2015 outburst when the source became too 
faint to be observed by Swift/BAT. The break occurred around MJD~$\sim\,57282$, which is very close 
to the transition from the super- to subcritical accretion regime, suggesting that these two events might be related. 
In this context, the break of the line decay could be associated with a transformation of the radiation-supported 
accretion column into an optically thin accretion mound, which should be present in the subcritical regime.

Moreover, by the 2016 outburst the cyclotron line energy  rebounded to the values found at 
the beginning of the 2015 outburst (Fig.~\ref{fig:drift}). If changes in the intrinsic magnetic 
field were responsible for the time evolution of the cyclotron line, as suggested by \citep{2016MNRAS.460L..99C}, it 
would imply an increase in the magnetospheric radius in 2016.  
Unfortunately, we could not estimate the magnetosphere size at later stages of the 2015 outburst or
for the 2016 outburst using the observed spin-up rates as suggested in \cite{2017MNRAS.466.2143D} 
because the spin-up rate could not be constrained due to the small time span between the two observations and 
insufficient counting statistics. 

However, the effective magnetospheric radius has been reported
to correlate with a high-frequency cutoff of a power density spectrum (PDS) of
nonperiodic flux fluctuations \citep{2009A&A...507.1211R,2014A&A...561A..96D},
which provides an alternative probe for the effective magnetosphere radius.
Indeed, the nonperiodic flux variability is believed to be due to accretion rate fluctuations throughout the accretion
disk on local Keplerian timescales \citep{1997MNRAS.292..679L}, so that the
high-frequency cutoff in the PDS is associated with the truncation of the accretion
disk by the magnetosphere, which is expected to occur close to the local
Keplerian frequency at the inner edge of the disk \citep{2009A&A...507.1211R,2014A&A...561A..96D}. 
A higher frequency of the cutoff thus corresponds to a smaller radius of the inner disk edge and the magnetosphere 
at a given accretion rate.

\begin{figure}[h]
\center{\includegraphics[width=1\linewidth]{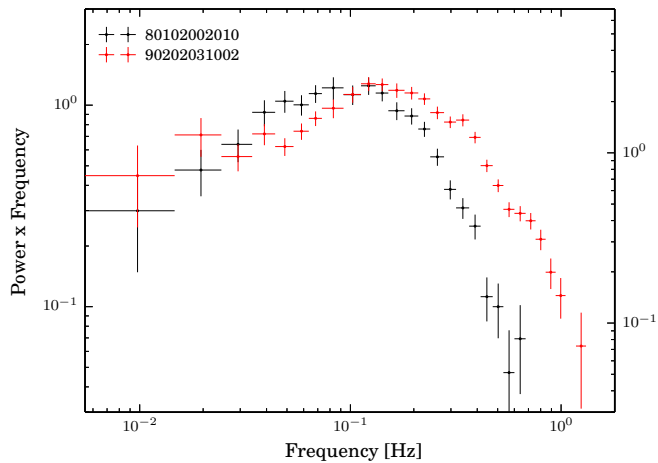}}
\caption{ NuSTAR power density spectra of V~0332+53 of the last 2015 observation (black crosses) 
and the first observation of the 2016 outburst (red crosses). 
The apparent excess at high frequencies is clearly seen in the spectrum of the 2016 observation, suggesting 
a decrease in the effective magnetospheric radius between the outbursts.
The power spectra were built with the \texttt{powspec} utility from HEASoft using the Leahy normalization 
\citep{1983ApJ...266..160L} with  white noise subtraction.}
\label{fig:powspec}
\end{figure}

To obtain the power density spectra, we use the same light curves, which 
were used to produce pulse profiles, as an input for the \texttt{powspec} utility from HEASoft. 
The result is present in Fig.~\ref{fig:powspec}, where the y-axis represents a power in terms of the Leahy 
normalization \citep{1983ApJ...266..160L} with a white noise subtraction (normalization=-1 
in \texttt{powspec}) multiplied by frequency. We also verified that the white noise power has the expected level and is not distorted 
by instrumental effects.
Figure~\ref{fig:powspec} shows  significantly
higher frequencies of the variability in 2016 compared to 2015 at similar luminosities, which should correspond to 
a smaller effective magnetospheric radius during the 2016 outburst. This contradicts the idea 
that changes in the intrinsic magnetic field are responsible for the time evolution of the line energy sustaining the argumentation from
\cite{2017MNRAS.466.2143D}.
Thus, the observed rebound of the line energy between the outbursts is most likely   
related to changes in the emission region configuration in this case as well. 

In particular, a smaller magnetosphere corresponds to a larger area of the polar region where the cyclotron line 
is believed to form because it depends on the effective magnetospheric radius $R_\textrm{m}$ as $1/R_\textrm{m}$.
Since the source was in the subcritical regime during the last two observations of the 2015 giant outburst
and during the 2016 outburst, the shock is no longer radiatively dominated and the infalling matter decelerates 
at some height either via Coulomb interactions 
\citep{2007A&A...465L..25S,2014A&A...572A.119S,2016A&A...590A..91S}
or in a collisionless shock above the surface, which is supposed to form 
due to  collective effects in the plasma 
\citep{1982ApJ...257..733L,2004AstL...30..309B}, 
leading to a positive $E_{\textrm{cyc}}/L_{\textrm{X}}$ correlation.

The CRSF is thought to arise mostly in the vicinity of the
collisionless shock whose height is proportional to the electron density in the emission
region \citep{1975ApJ...198..671S}.
This model was successfully applied to interpret the $E_{\textrm{cyc}}/L_{\textrm{X}}$ correlations in Cep~X-4
\citep{2017A&A...601A.126V} and GX~304-1 \citep{2017MNRAS.466.2752R} as well as
the change in a spectral hardness in Cep X-4 \citep{2017A&A...601A.126V}. 
As discussed above, the PDSs point to a larger polar cap area in the 2016 observations at a luminosity 
comparable to that found during the two last observations in 2015, implying a lower electron density in the emission region in 2016.
This is expected to reduce the height of the collisionless shock and thus to increase the
CRSF energy as observed.

An alternative model suggested by \cite{2015MNRAS.454.2714M} to explain a 
$E_{\textrm{cyc}}/L_{\textrm{X}}$ 
correlation in subcritical sources relates the observed changes in
cyclotron line energy to the Doppler effect associated with the bulk motion of
the accreting flow in the vicinity of the neutron star surface. 
In this model the observed CRSF appears
red-shifted by an amount proportional to the accretion flow velocity, which is
expected to decrease with higher accretion rate, so that the line energy appears to be
correlated with luminosity. Since the local electron density decreases with the 
increasing polar cap area at a given mass accretion rate, the flow velocity in the vicinity of the hot spot is expected
to increase according to this model, i.e.,  for V~0332+53 one could expect a higher flow velocity and 
a lower line energy in the 2016 outburst, which does not agree with observations.

For V~0332+53, the observed rebound of the cyclotron line energy by the 2016
outburst is therefore explained rather with the collisionless shock model. 
We note, however, that the Doppler effect must still play a role also in this case, and
should be taken into account. We conclude, therefore, that the  results presented here
can be very useful for the further development and verification of these models.

\section{Conclusions}
Using NuSTAR observations of the Be-transient X-ray pulsar V~0332+53
carried out during a giant outburst in 2015, and a subsequent ordinary
outburst in 2016, we investigated for the first time the dependence of the CRSF centroid energy on luminosity
both on long (months) and short (seconds) timescales using the  pulse-to-pulse
analysis technique.

We found that the dependence is essentially the same on both timescales and confirmed
with high significance the reported transition from an anti-correlation to a correlation
at $L_{\textrm{cr}} = (2.1 \pm  0.4) \times 10^{37}$ erg/s, which is likely associated 
with the transition between the super- and subcritical accretion regimes defined according 
to the presence/absence of the radiation-supported accretion column. 

We also found that the time-linear decay during the 2015 outburst is likely to 
break up at the moment of the transition between the two types of the $E_{\textrm{cyc}}/L_{\textrm{X}}$ 
correlation, which could be associated with changes in the accretion structure above the polar cap caused by 
switching between the super- and subcritical accretion regimes.

The line energy then rebounded between the 2015 and 2016 outbursts to values observed
at the beginning of the giant outburst. We argue that this change is also likely
related to a change in the geometrical configuration of the emission region.
This conclusion is supported by the analysis of power density spectra of  nonperiodic flux fluctuations,
which exhibits  variability at a significantly higher frequency in the 2016 outburst compared with the 
decay phase of the 2015 outburst. 
We interpret this as  evidence of a smaller truncation radius of the accretion disk, and as
a result, a smaller magnetosphere in 2016. 
We briefly discuss how such a change is expected to affect 
the observed line energy in the framework of the model proposed by 
\cite{2015MNRAS.454.2714M} and the model with a collisionless shock 
\citep{2017MNRAS.466.2752R,2017A&A...601A.126V}.
We conclude that the latter model more readily explains the observations.

We note that a long-term evolution has been reported before for Her~X$-$1 
\citep{2014A&A...572A.119S,2016A&A...590A..91S} and Vela X$-$1 \citep{2016MNRAS.463..185L}. 
In the case of V~0332+53, we argue that the observed time evolution of the CRSF energy is 
driven by changes in the geometry of the emission region.

\begin{acknowledgements}
The research is supported by the joint DFG grant KL~2734/2-1 and Wi~1860~11-1. 
We thank the anonymous referee for the useful comments and suggestions that substantially improved the manuscript.
The research used the data obtained from HEASARC Online Service provided by the NASA/GSFC.
\end{acknowledgements}

\bibliographystyle{aa}
\bibliography{ref}

\begin{thebibliography}{28}
\expandafter\ifx\csname natexlab\endcsname\relax\def\natexlab#1{#1}\fi

\bibitem[{{Basko} \& {Sunyaev}(1976)}]{1976MNRAS.175..395B}
{Basko}, M.~M. \& {Sunyaev}, R.~A. 1976, \mnras, 175, 395

\bibitem[{{Bykov} \& {Krasilshchikov}(2004)}]{2004AstL...30..309B}
{Bykov}, A.~M. \& {Krasilshchikov}, A.~M. 2004, Astronomy Letters, 30, 309

\bibitem[{{Cusumano} {et~al.}(2016){Cusumano}, {La Parola}, {D'A{\`i}},
  {Segreto}, {Tagliaferri}, {Barthelmy}, \& {Gehrels}}]{2016MNRAS.460L..99C}
{Cusumano}, G., {La Parola}, V., {D'A{\`i}}, A., {et~al.} 2016, \mnras, 460,
  L99

\bibitem[{{Doroshenko} {et~al.}(2014){Doroshenko}, {Santangelo}, {Doroshenko},
  {Caballero}, {Tsygankov}, \& {Rothschild}}]{2014A&A...561A..96D}
{Doroshenko}, V., {Santangelo}, A., {Doroshenko}, R., {et~al.} 2014, \aap, 561,
  A96

\bibitem[{{Doroshenko} {et~al.}(2017){Doroshenko}, {Tsygankov}, {Mushtukov},
  {Lutovinov}, {Santangelo}, {Suleimanov}, \& {Poutanen}}]{2017MNRAS.466.2143D}
{Doroshenko}, V., {Tsygankov}, S.~S., {Mushtukov}, A.~A., {et~al.} 2017,
  \mnras, 466, 2143

\bibitem[{{F{\"u}rst} {et~al.}(2014){F{\"u}rst}, {Pottschmidt}, {Wilms},
  {Tomsick}, {Bachetti}, {Boggs}, {Christensen}, {Craig}, {Grefenstette},
  {Hailey}, {Harrison}, {Madsen}, {Miller}, {Stern}, {Walton}, \&
  {Zhang}}]{2014ApJ...780..133F}
{F{\"u}rst}, F., {Pottschmidt}, K., {Wilms}, J., {et~al.} 2014, \apj, 780, 133

\bibitem[{{Heindl} {et~al.}(2004){Heindl}, {Rothschild}, {Coburn}, {Staubert},
  {Wilms}, {Kreykenbohm}, \& {Kretschmar}}]{2004AIPC..714..323H}
{Heindl}, W.~A., {Rothschild}, R.~E., {Coburn}, W., {et~al.} 2004, in American
  Institute of Physics Conference Series, Vol. 714, X-ray Timing 2003: Rossi
  and Beyond, ed. P.~{Kaaret}, F.~K. {Lamb}, \& J.~H. {Swank}, 323--330

\bibitem[{{Klochkov} {et~al.}(2011){Klochkov}, {Staubert}, {Santangelo},
  {Rothschild}, \& {Ferrigno}}]{2011A&A...532A.126K}
{Klochkov}, D., {Staubert}, R., {Santangelo}, A., {Rothschild}, R.~E., \&
  {Ferrigno}, C. 2011, \aap, 532, A126

\bibitem[{{Kreykenbohm} {et~al.}(2005){Kreykenbohm}, {Mowlavi}, {Produit},
  {Soldi}, {Walter}, {Dubath}, {Lubi{\'n}ski}, {T{\"u}rler}, {Coburn},
  {Santangelo}, {Rothschild}, \& {Staubert}}]{2005A&A...433L..45K}
{Kreykenbohm}, I., {Mowlavi}, N., {Produit}, N., {et~al.} 2005, \aap, 433, L45

\bibitem[{{La Parola} {et~al.}(2016){La Parola}, {Cusumano}, {Segreto}, \&
  {D'A{\`i}}}]{2016MNRAS.463..185L}
{La Parola}, V., {Cusumano}, G., {Segreto}, A., \& {D'A{\`i}}, A. 2016, \mnras,
  463, 185

\bibitem[{{Langer} \& {Rappaport}(1982)}]{1982ApJ...257..733L}
{Langer}, S.~H. \& {Rappaport}, S. 1982, \apj, 257, 733

\bibitem[{{Leahy} {et~al.}(1983){Leahy}, {Darbro}, {Elsner}, {Weisskopf},
  {Kahn}, {Sutherland}, \& {Grindlay}}]{1983ApJ...266..160L}
{Leahy}, D.~A., {Darbro}, W., {Elsner}, R.~F., {et~al.} 1983, \apj, 266, 160

\bibitem[{{Lyubarskii}(1997)}]{1997MNRAS.292..679L}
{Lyubarskii}, Y.~E. 1997, \mnras, 292, 679

\bibitem[{{Malacaria} {et~al.}(2015){Malacaria}, {Klochkov}, {Santangelo}, \&
  {Staubert}}]{2015A&A...581A.121M}
{Malacaria}, C., {Klochkov}, D., {Santangelo}, A., \& {Staubert}, R. 2015,
  \aap, 581, A121

\bibitem[{{Mushtukov} {et~al.}(2015){Mushtukov}, {Tsygankov}, {Serber},
  {Suleimanov}, \& {Poutanen}}]{2015MNRAS.454.2714M}
{Mushtukov}, A.~A., {Tsygankov}, S.~S., {Serber}, A.~V., {Suleimanov}, V.~F.,
  \& {Poutanen}, J. 2015, \mnras, 454, 2714

\bibitem[{{Poutanen} {et~al.}(2013){Poutanen}, {Mushtukov}, {Suleimanov},
  {Tsygankov}, {Nagirner}, {Doroshenko}, \& {Lutovinov}}]{2013ApJ...777..115P}
{Poutanen}, J., {Mushtukov}, A.~A., {Suleimanov}, V.~F., {et~al.} 2013, \apj,
  777, 115

\bibitem[{{Protassov} {et~al.}(2002){Protassov}, {van Dyk}, {Connors},
  {Kashyap}, \& {Siemiginowska}}]{2002ApJ...571..545P}
{Protassov}, R., {van Dyk}, D.~A., {Connors}, A., {Kashyap}, V.~L., \&
  {Siemiginowska}, A. 2002, \apj, 571, 545

\bibitem[{{Revnivtsev} {et~al.}(2009){Revnivtsev}, {Churazov}, {Postnov}, \&
  {Tsygankov}}]{2009A&A...507.1211R}
{Revnivtsev}, M., {Churazov}, E., {Postnov}, K., \& {Tsygankov}, S. 2009, \aap,
  507, 1211

\bibitem[{{Rothschild} {et~al.}(2017){Rothschild}, {K{\"u}hnel}, {Pottschmidt},
  {Hemphill}, {Postnov}, {Gornostaev}, {Shakura}, {F{\"u}rst}, {Wilms},
  {Staubert}, \& {Klochkov}}]{2017MNRAS.466.2752R}
{Rothschild}, R.~E., {K{\"u}hnel}, M., {Pottschmidt}, K., {et~al.} 2017,
  \mnras, 466, 2752

\bibitem[{{Shapiro} \& {Salpeter}(1975)}]{1975ApJ...198..671S}
{Shapiro}, S.~L. \& {Salpeter}, E.~E. 1975, \apj, 198, 671

\bibitem[{{Staubert} {et~al.}(2016){Staubert}, {Klochkov}, {Vybornov}, {Wilms},
  \& {Harrison}}]{2016A&A...590A..91S}
{Staubert}, R., {Klochkov}, D., {Vybornov}, V., {Wilms}, J., \& {Harrison},
  F.~A. 2016, \aap, 590, A91

\bibitem[{{Staubert} {et~al.}(2014){Staubert}, {Klochkov}, {Wilms}, {Postnov},
  {Shakura}, {Rothschild}, {F{\"u}rst}, \& {Harrison}}]{2014A&A...572A.119S}
{Staubert}, R., {Klochkov}, D., {Wilms}, J., {et~al.} 2014, \aap, 572, A119

\bibitem[{{Staubert} {et~al.}(2007){Staubert}, {Shakura}, {Postnov}, {Wilms},
  {Rothschild}, {Coburn}, {Rodina}, \& {Klochkov}}]{2007A&A...465L..25S}
{Staubert}, R., {Shakura}, N.~I., {Postnov}, K., {et~al.} 2007, \aap, 465, L25

\bibitem[{{Titarchuk}(1994)}]{1994ApJ...434..570T}
{Titarchuk}, L. 1994, \apj, 434, 570

\bibitem[{{Trümper} {et~al.}(1978){Trümper}, {Pietsch}, {Reppin}, {Voges},
  {Staubert}, \& {Kendziorra}}]{1978ApJ...219L.105T}
{Trümper}, J., {Pietsch}, W., {Reppin}, C., {et~al.} 1978, \apjl, 219, L105

\bibitem[{{Tsygankov} {et~al.}(2006){Tsygankov}, {Lutovinov}, {Churazov}, \&
  {Sunyaev}}]{2006MNRAS.371...19T}
{Tsygankov}, S.~S., {Lutovinov}, A.~A., {Churazov}, E.~M., \& {Sunyaev}, R.~A.
  2006, \mnras, 371, 19

\bibitem[{{Vybornov} {et~al.}(2017){Vybornov}, {Klochkov}, {Gornostaev},
  {Postnov}, {Sokolova-Lapa}, {Staubert}, {Pottschmidt}, \&
  {Santangelo}}]{2017A&A...601A.126V}
{Vybornov}, V., {Klochkov}, D., {Gornostaev}, M., {et~al.} 2017, \aap, 601,
  A126

\bibitem[{{Yamamoto} {et~al.}(2011){Yamamoto}, {Sugizaki}, {Mihara},
  {Nakajima}, {Yamaoka}, {Matsuoka}, {Morii}, \&
  {Makishima}}]{2011PASJ...63S.751Y}
{Yamamoto}, T., {Sugizaki}, M., {Mihara}, T., {et~al.} 2011, \pasj, 63, S751

\end{thebibliography}

\end{document}